\pdfoutput=1

\documentclass[usegraphicx,usenatbib]{mn2e}

\usepackage{amssymb,bm,color,charter}

\usepackage{ulem}

\font\tenbg=cmmib10 at 10pt

\def \rvecphi{{\hbox{\tenbg\char'036}}}

\def\lesssim{\mathrel{\hbox{\rlap{\hbox{\lower4pt\hbox{$\sim$}}}\hbox{$<$}}}}
\def\gtrsim{\mathrel{\hbox{\rlap{\hbox{\lower4pt\hbox{$\sim$}}}\hbox{$>$}}}} 

\title[Ring Formation]{Ring Formation from an Oscillating Black Hole}

\author[Lovelace \& Kornreich]{R.V.E. Lovelace,$^{1}$\thanks{E-mail: RVL1@astro.cornell.edu}
D.A. Kornreich,$^2$\thanks{David.Kornreich@humboldt.edu}\\
$^1$ Departments of Astronomy and Applied and Engineering
Physics, Cornell University, Ithaca, NY 14853, USA\\
$^2$ Department of Physics and Astronomy, 
Humboldt State University, Arcata, CA 95521, USA}

\begin{document}
\maketitle 
\label{firstpage}

\begin{abstract}

Massive black hole (BH) mergers can result in the merger remnant receiving a ``kick,'' of order 200 km~s$^{-1}$ or more, which will cause the remnant to oscillate about the galaxy centre. 
   Here we analyze the case where the BH oscillates through
the galaxy centre   perpendicular or parallel to the plane of the galaxy  for
a model galaxy consisting of an
exponential disk, a Plummer model bulge, and an isothermal 
dark matter halo.
    For the perpendicular motion we find 
that there is a strong resonant  forcing
of the disk radial motion near but somewhat
less than the ``resonant radii'' $r_R$ where the BH oscillation frequency is equal one-half, one-fourth, ($1/6$, etc.) of the radial
epicyclic frequency in the plane of the disk.
    Near the resonant radii
there can be a strong enhancement of the radial flow and
disk density which can lead to shock formation. 
In turn the shock may trigger the
formation of a ring of stars near $r_R$.
    As an example, for a BH mass of $10^8~M_\odot$ and a kick
velocity of $150$ km s$^{-1}$,    
we find that the resonant radii lie between $0.2$ and $1$ kpc.
    For BH motion parallel to the plane of the galaxy
we find that the BH leaves behind it a supersonic wake
where star formation may be triggered.   The shape  of the
wake is calculated as well as the slow-down time of the BH.
  The differential rotation of the disk stretches the wake into
ring-like segments.

\end{abstract}


\section{Introduction}

   Recent breakthroughs in numerical General Relativity have led to predictions of large recoil velocities of merged binary supermassive black-holes (BH) as the binary radiates away linear momentum as gravitational waves during the final stages of merger (Gonz\'alez et al. 2007;  Campanelli et al. 2007;  Lousto, Campanelli, \& Zlochower 2009). 
     Typical kick velocities are of order $\sim 200$~km~s$^{-1}$ 
 (Bogdanovi\'c, Reynolds, \& Coleman 2007). 
  One possible result of these mergers is to cause the resulting remnant black hole (BH) to oscillate through the stellar disk on timescales on the order of Gyr (Kornreich \& Lovelace 2008, hereafter KL08; Blecha \& Loeb 2008; Fujita 2009).
If the merged BH receives an impulse from the merger, some of that motion will be transmitted to the galaxy by dynamical friction. 
   It is clearly of interest to  know whether 
the motion of the BH through the disk 
generates observable changes in morphology and
dynamics in the galaxy.
       Binary black-holes have already
been observed as a double nucleus in a quasar (e.g., Decarli et al. 2009).
     Further, Comerford et al. (2009) show that 
BH binaries resulting from recent mergers are 
observable in the spectra of as many as $40\%$ of Seyfert 2 galaxies. The possible observable effects of free and ``wandering''  BH merger remnants on disk
galaxies, however, have not yet been fully analyzed.     
de la Fuente Marcos \& de la Fuente Marcos (2008) discuss the formation of
stars in the wakes of runaway black holes ejected from the host galaxy.
      Here we consider more slowly moving BHs which remain bound
to the host galaxy.

      This work first analyzes the case where the
ejected BH oscillates with angular frequency $\Omega_{bhz}$
through the galaxy center in a direction normal to the disk.     
   Later, we discuss the case where the BH
is ejected parallel to the plane of the galaxy.
   Sufficiently large gas accretion by the binary BH 
system is predicted to drive the orbital and BH spins into alignment
normal to the plane of the galaxy (Bogdanovi\'c et al. 2007).   
However, the amount is uncertain due to the uncertainty
in  the ratio of the two viscosity coefficients for the $(r,z)$ and
$(r,\phi)$ motion in the disk (e.g., Natarajan \& Armitage 1999).
     This alignment  favors BH ejection 
in the plane of the galaxy (Campanelli et al. 2007).
     The merging process preceding the ejection is
assumed to be sufficiently slow that the  host galaxy has
returned to an axisymmetric equilibrium state.   
   The BH oscillations are observed in simulations 
to persist for many periods (KL08; Blecha \& Loeb 2008; Fujita 2009).     
    Under these conditions we find 
that there is a strong resonant  forcing
of the disk radial motion near radii $r_R$ where $2n\Omega_{bhz}=
\Omega_r(r_R)$, with $n=1,~2,..$, where $\Omega_r$ is 
the radial epicyclic frequency.   
   Near $r_R$ there can be a strong enhancement of the
density in  a ring of the disk
and shock formation which may 
trigger the formation of a ring of stars.
      We note that models for the formation of observed
``ring galaxies'' assume the (single) passage of one
galaxy nearly through the center of another
(Theys \& Spiegel 1976, 1977; Lynds \& Toomre 1976).
   The gravitational interaction of the two galaxies causes
the formation of a pronounced ring(s) of enhanced
density  (in one or both of the galaxies) where star
formation is expected to be enhanced.   

   For the case of BH ejection in the plane of the
galaxy we find that the BH leaves behind it a 
supersonic wake where star formation may be
triggered.  The shape of this wake is calculated and
the slow-down time of the BH is estimated.  
      The differential rotation of the disk stretches the wake into
ring-like segments.

   Section 2.1 describes the gravitational potential 
of the equilibrium disk galaxy at the
time of the BH merger.    This includes an
exponential disk of stars and gas,  a Plummer model the bulge,
and an isothermal dark matter halo. 
   Section 2.2 derives the dynamical equations 
of the system as driven by the radial force 
due to a BH oscillating perpendicular to the disk.
    The frequencies and amplitudes of the BH oscillations are 
taken from the   $N$--body simulations of KL08, 
    Section 2.3 discusses the forced oscillations of the
gas disk, and \S 2.4 the possible parametric instability
of the disk.    
    Section 3 discusses  the response of
a galaxy disk to a BH ejected in the
plane of the galaxy.
   Section 4 gives the conclusions of this work.

\section{Theory}

  \subsection{Equilibrium}

The equilibrium galaxy is assumed to be 
axisymmetric and to consist of a thin disk of
stars and gas and a spheroidal 
distributions consisting of
a bulge component and a halo of
dark matter.   The model is the same
as that used in the simulations of KH08.
    The model is similar to that of
Fujita (2008, 2009).    
We use an inertial cylindrical $(r,\phi,z)$
and Cartesian $(x,y,z)$ coordinate systems with the
disk and halo equatorial planes in the $z=0$ plane.  
 The
total gravitational potential is written as
\begin{equation}
\Phi(r,z)=\Phi_{d} +\Phi_b+\Phi_h~, 
\end{equation}
where $\Phi_{d}$ is the potential due to the 
disk, $\Phi_b$ is due to the bulge,
and $\Phi_h$ is that for the
halo.  
  The galaxy may have a central massive
black hole of mass $M_{bh}$ in which
case a term $\Phi_{bh}=-GM_{bh}/\sqrt{r^2+z^2}$ is
added to the right-hand side of (1).
 The particle orbits in the equilibrium disk
are approximately circular with angular
rotation rate $\Omega(r)$, where
\begin{equation}
\Omega^2(r)={1\over r}
{{\partial \Phi}\over{\partial r}}\bigg|_{z=0} =
\Omega_{d}^2+\Omega_b^2+\Omega_h^2~.
\end{equation}
The equilibrium disk velocity is 
${\bf v} = r \Omega(r) \hat{\rvecphi~}.$
 A central black hole is  accounted for
by adding the term $\Omega_{bh}^2=GM_{bh}/r^3$
to the right-hand side of (2).

   The surface mass density of the (optical) disk 
is taken to be $\Sigma_d =\Sigma_{d0} {\rm exp}
(-r/r_d)$ with $\Sigma_{d0}$ and $r_d$  
constants and $M_d=2\pi r_d^2\Sigma_{d0}$ the
total disk mass.  The potential
due
to this disk matter is
\begin{equation}
\Phi_d(r,0)=-~{G M_d \over r_d}
R[I_0(R)K_1(R)
-I_1(R)K_0(R)] ~,
\end{equation}
and the corresponding angular velocity is
\begin{equation}
\Omega_d^2={1\over 2}{G M_d \over r_d^3}
\left[I_0(R)K_0(R)
-I_1(R) K_1(R) \right]~,
\end{equation}
where $R\equiv r/(2r_d)$
and the $I's$ and $K's$ are
the usual modified Bessel 
functions (Freeman 1970;  Binney \& Tremaine 1987, p.77).
  We consider the mass of the disk stars and gas is
$M_d = 2.8 \times 10^{10}{\rm M}_\odot$ 
and $r_d = 3.5$ kpc following KL08.
  For these values, $v_d \equiv \sqrt{GM_d/r_d}
= 186$ km/s.

\begin{figure}
\centering
\includegraphics[scale=0.45]{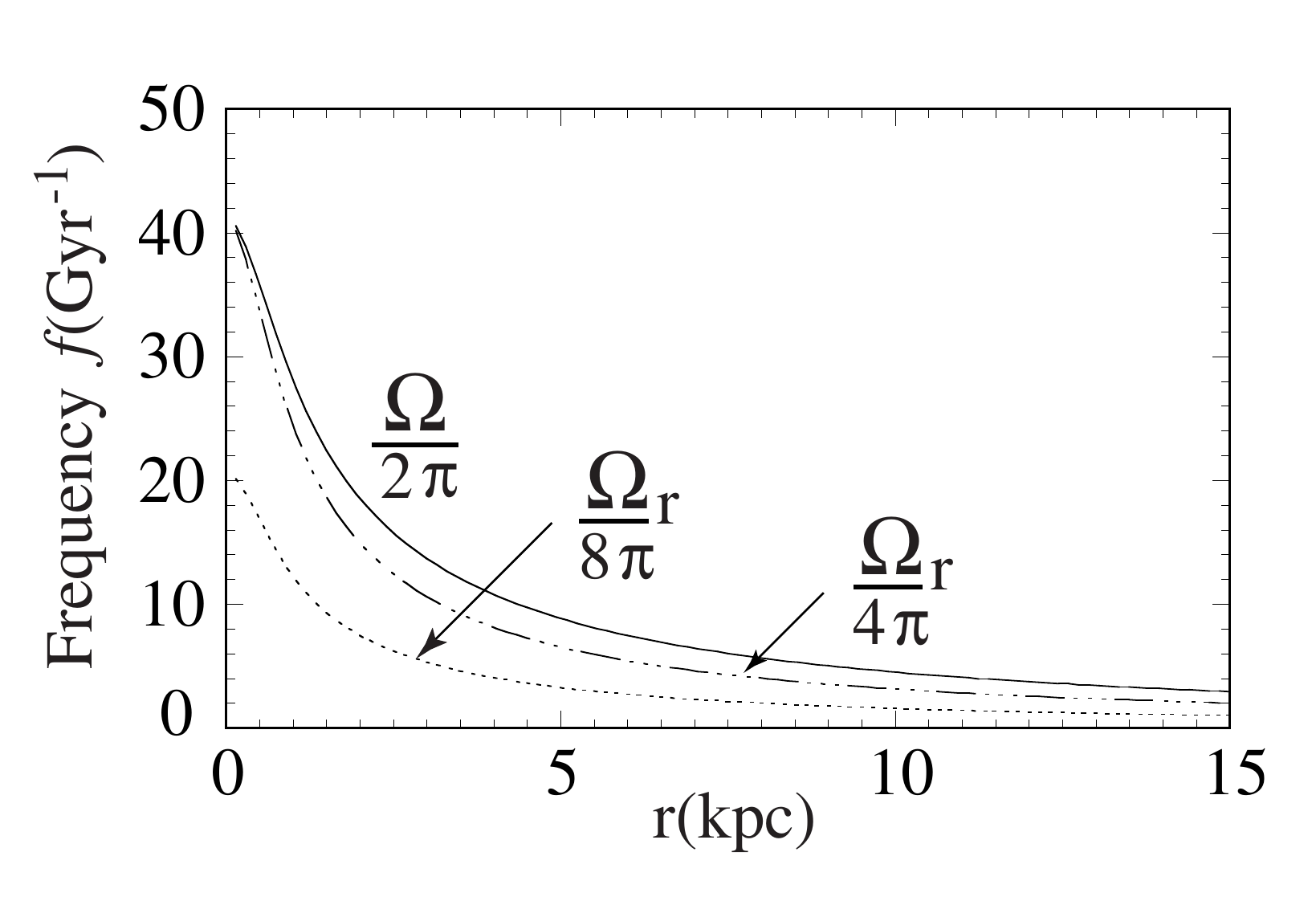}
\caption{Rotation frequency of the disk $f=\Omega/2\pi$,
 one-half the radial epicyclic frequency $f_r/2=\Omega_r/4\pi$,
and one-fourth this value, $f_r/4$,
for the galaxy model described in \S 2.2.
}
\end{figure}

  The potential due to the bulge component
is taken as a Plummer (1915) model
\begin{equation}
\Phi_b =-~{G M_b \over (r_b^2+r^2+z^2)^{1/2}}~,
\end{equation}
where $M_b$ is the mass of the bulge and $r_b$
is its characteristic radius (Binney \& Tremaine
1987, p.42).  This component is similar
to the spherical component considered
by Fujita (2008, 2009).
We have
\begin{equation}
\Omega_b^2 = { G M_b \over
(r_b^2 + r^2)^{3/2}}~.
\end{equation}
   We take
$M_b = 10^{10} {\rm M}_{\odot}$ and $r_b =
1$ kpc so that $v_b \equiv \sqrt{GM_b/r_b}
= 208~ {\rm km/s}$ again following KL08.

      The  dark matter halo  is assumed
to have the isothermal distribution with
the potential
\begin{equation}
\Phi_h = {1\over 2}v_{h}^2 ~
\ln(r_{c}^2+{ r}^2+z^2)~,
\end{equation}
where  $r_{c}$  is the core radius of the halo,
 and $v_{h}$  is the circular 
velocity at distances larger than $r_c$.
This potential is assumed to apply out to
a large distance say $50$ kpc.
   Equation (7) implies
\begin{equation}
\Omega_h^2 = {v_{h}^2 \over r_{h}^2+{ r}^2}~.
\end{equation}
Representative values are $v_{h} = 250$ km/s 
and $r_{c}= 2$ kpc.  
    The total dynamical mass of the galaxy model is 
$M_{\rm tot}\approx 6\times 10^{11}$.
    We do not attempt to model the dark matter distribution
at small radii where 
 $\Lambda$CDM simulations predict a
 cusp with $\rho_{dm} \propto r^{-1}$ 
(Navarro, Frenk, \& White 1996).   
     Observed rotation curves of spiral galaxies
are however well fit by profiles $\rho_{dm}
\propto (r_B+r)^{-1}$ with $r_B$ is larger
than $5$ kpc (Salucci 2001;  Salucci \& Burkert 2000;
Burkert 1995).     

   Figure 1 shows
an illustrative rotation curve $\Omega(r)/2\pi$
{\it not} including the contribution of the BH.
The contribution of the BH for $M_{bh}=10^8M_\odot$ 
is not significant for $r>0.2 $ kpc.

\subsection{Oscillating Black Hole}

We first consider the case where the BH of mass $M_{bh}$
is ejected vertically and oscillates 
about the midplane of the galaxy,
\begin{equation}
z_{bh}= z_{m}\big[ \sin(\Omega_{bhz} t)
+c_3\sin(3\Omega_{bhz}t)+..\big]~,
\end{equation}
where $\Omega_{bhz} =2\pi f_{bhz}$ is
the angular frequency of the vertical oscillation, and
$c_3,~c_5,..$ account for the  fact that the motion is 
not in general simple harmonic, but the effective
potential is an even function of $z$.  
However, for
$z_m \lesssim 1$ kpc, the magnitudes the $c's$ is
small compared with unity (KL08), and we neglect them.
    Figure 2 shows the dependence of $f_{bhz}$
on the initial kick velocity  of the BH from KL08.
    We assume that the damping time-scale
of the oscillations due to dynamical friction is
significantly   longer than the oscillation period as found
by KL08,  Blecha \& Loeb (2008), and Fujita (2009).

\begin{figure}
\centering
\includegraphics[scale=0.45]{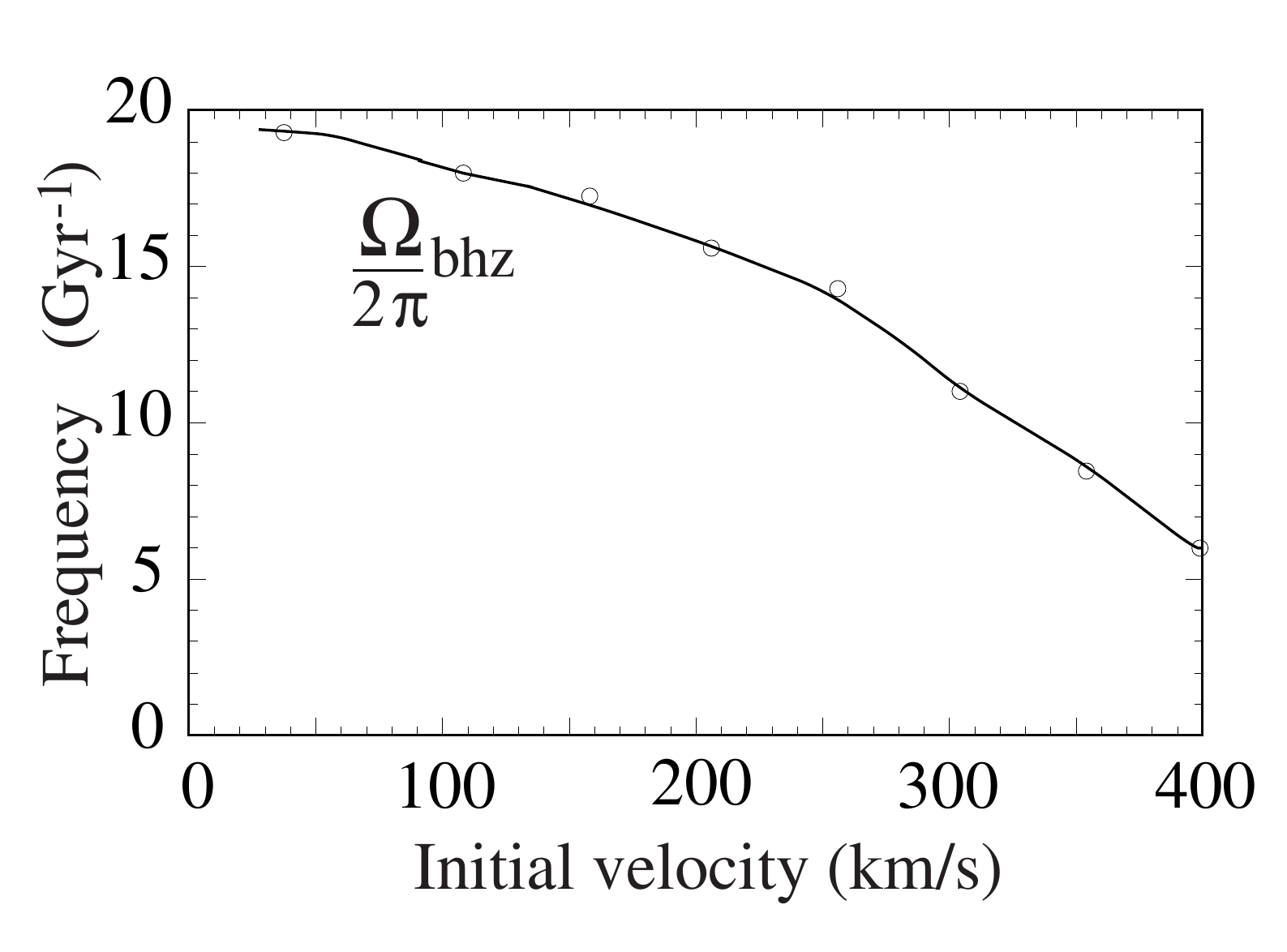}
\caption{Vertical oscillation frequency $f_{bhz}=\Omega_{bhz}/2\pi$
as a function of the initial velocity of the BH from KL08
for $M_{bh}=10^8M_\odot$.   The line through the
points is simply a smooth fit.}
\end{figure}

\begin{figure}
\centering
\includegraphics[scale=0.45]{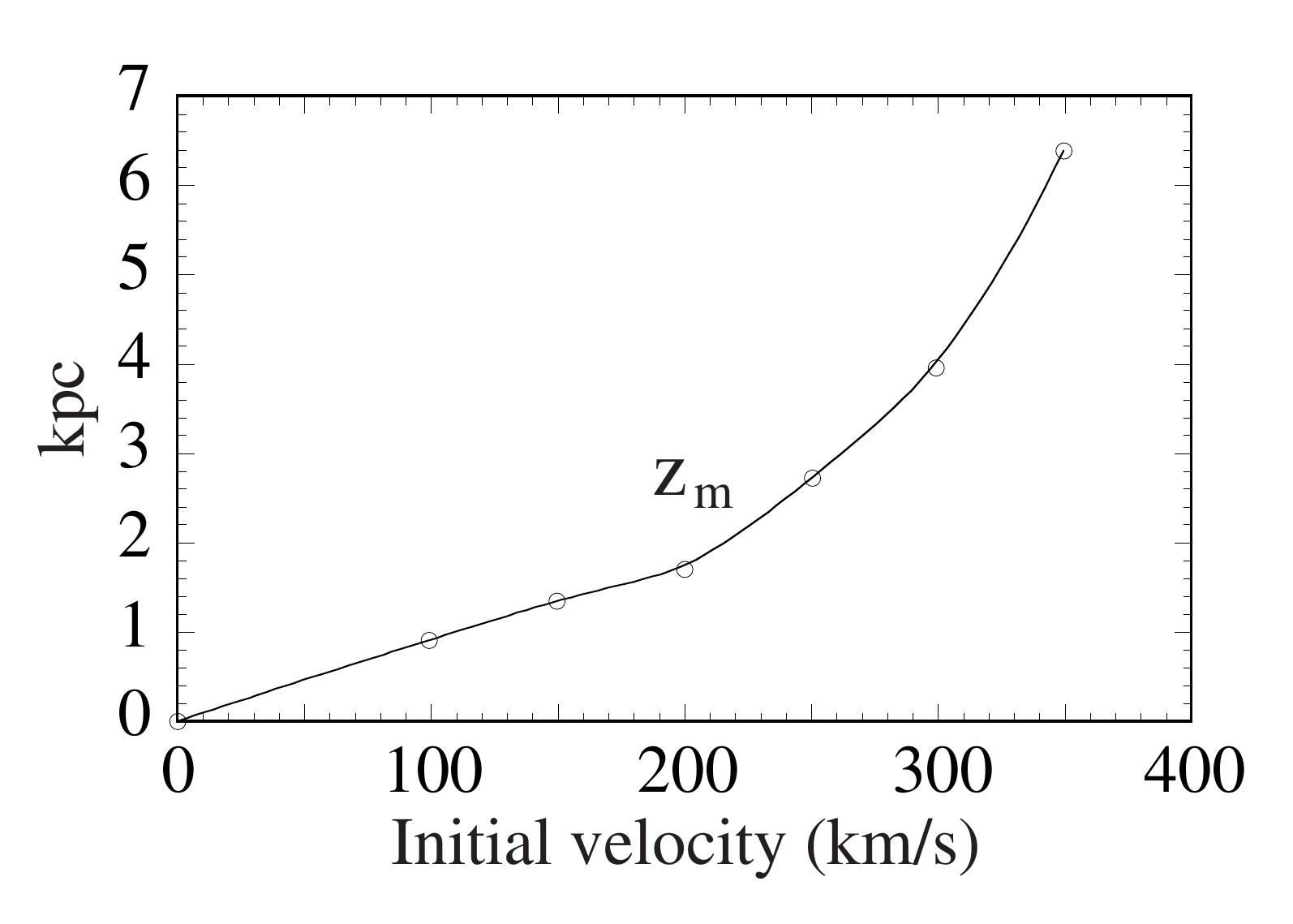}
\caption{Dependence of the vertical
amplitude of the BH motion $z_{bh0}$ on the initial
velocity from KL08 for $M_{bh}=10^8M_\odot$. 
The line through the points is simply a smooth fit. }
\end{figure}

\begin{figure}
\centering
\includegraphics[scale=0.45]{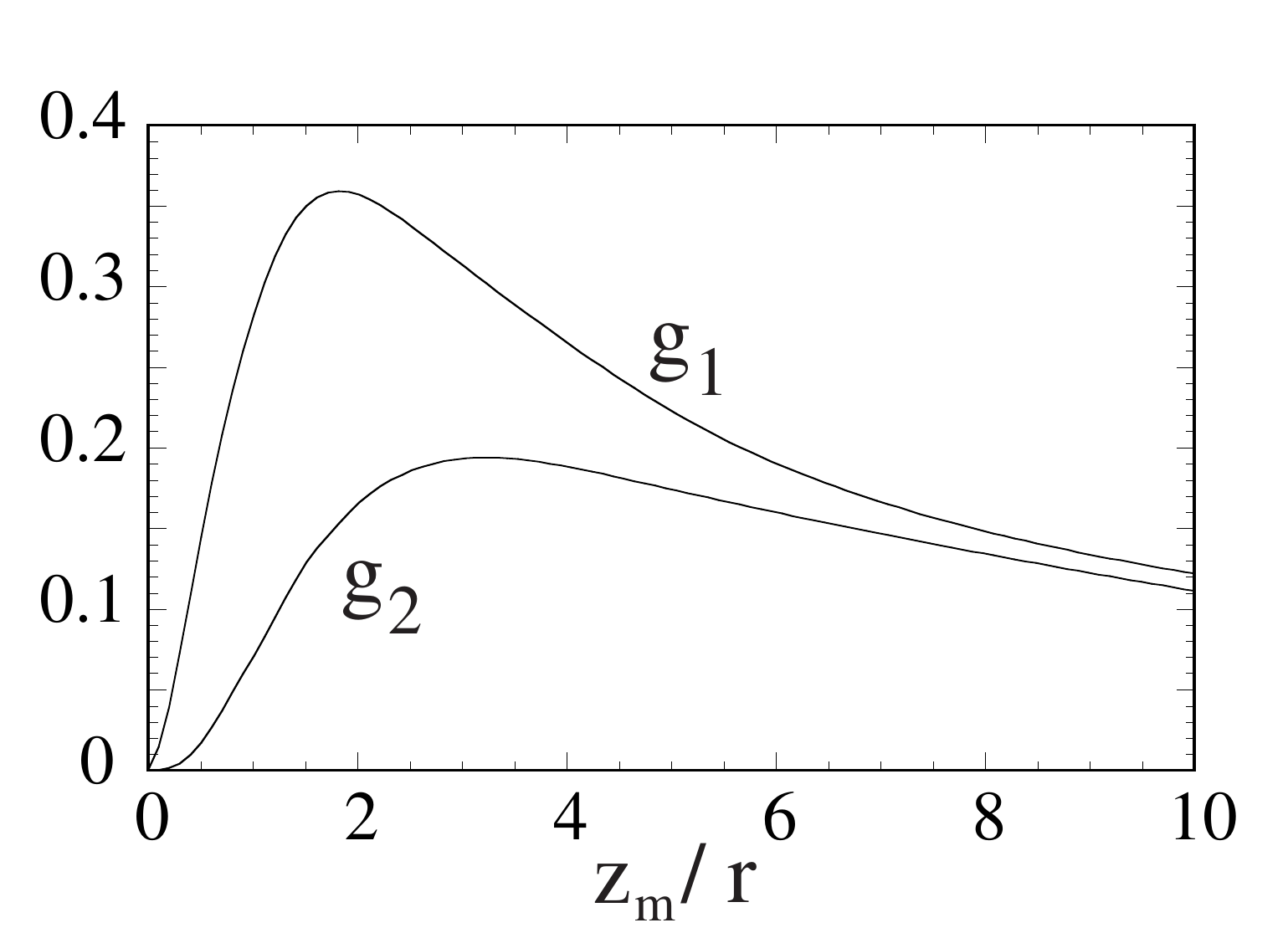}
\caption{Dimensionless coefficients dependent
only on $z_m/r_0$ as  defined in
equation (11) and obtained numerically.  }
\end{figure}
   
The oscillating BH gives a time-dependent radial force or
acceleration on the disk matter at a radius $r$,
\begin{equation}
a_r(r,t) = -{GM_{bh} r \over [r^2 + z_{bh}^2(t)]^{3/2}}
= - \Omega_{bh}^2~ r ~ {\cal F}~,
\end{equation}
where $\Omega_{bh} \equiv \sqrt{G M_{bh}/r^3}$, and
\begin{eqnarray}
{\cal F}& =&\left[1+\left({z_{bh} \over r}\right)^2\right]^{-3/2}
\nonumber\\
&=&g_0+
g_1\cos(2\Omega_{bhz} t) + g_2 \cos(4\Omega_{bhz} t)+...~.
\end{eqnarray}
Here, $g_0$, $g_1$, and $g_2$ are dimensionless functions of $z_{m}/r$ as  shown in Figure 4.    The $g_0$ term acts to modify the disk equilibrium
mainly at distances $\lesssim 0.2$ kpc,
and it is not considered further.

     We model the disk as an axisymmetric thin fluid
layer  acted on by {\it only} the
radial   force of the BH ($a_r$).  
    The disk consists of gas and stars so that its {\it linear} response
to the black hole should be treated as two
gravitationally coupled fluids - gas
and stars - with different surface densities ($\Sigma_g$ 
and $\Sigma_*$) and different `sound' speeds
($c_g$ and $c_*$) following the approach of Lin \& Shu (1970).  
     In our galaxy $\Sigma_g \ll \Sigma_*$ and 
$c_g \ll c_*$,  The Toomre (1964) stability factor for
axisymmetric perturbations 
of the gas  $Q_g \propto c_g/\Sigma_g$ is
significantly less than that for the stars $Q_* \propto c_*/\Sigma_*$.
   A number of studies (e.g., 
Jog \& Soloman 1984; Wang \& Silk 1994;  Rafikov 2001) ) 
point out that the disk stability is 
determined mainly by the small-$Q$ (less
stable) component  of the disk.  
For this reason we  analyze the
response of the gaseous component of the disk
and let $\Sigma = \Sigma_g$.
    We have
\begin{eqnarray}
{d u_r \over dt }& =& - {\nabla P \over \Sigma}
+{u_\phi^2 \over r} -{ \partial \Phi \over \partial r}+a_r(r,t)~,   
\nonumber \\
{d (r u_\phi) \over dt}& =&0~,
\end{eqnarray}
where $d/dt = \partial/\partial t +u_r(\partial/\partial r)$,  $\Sigma(r,t)$
is the surface mass-density,  and $P(r,t)$ is the height-integrated
pressure.   
    The disk mass is conserved so that
\begin{equation}
{\partial \Sigma\over \partial t}+{1\over r}
{\partial (r u_r \Sigma) \over \partial r}=0~.
\end{equation}  
There is also an equation for the gravitational
potential.

   We linearize equations (12) and (13) by letting
$u_r = 0 +\delta u_r$, $u_\phi =r \Omega  +\delta u_\phi$,
$P= P_0 +\delta P$,  $\Sigma=\Sigma_0+\delta \Sigma$, and
$\Phi = \Phi_0 +\delta \Phi$.   The disk equilibrium
has $0=-\Sigma_0^{-1}\partial P_0/\partial r +\Omega^2 r
-\partial \Phi_0/\partial r$.  Equations (12) then give
\begin{eqnarray}
{d\delta u_r \over dt}&=&{\delta \Sigma \over \Sigma_0^2}
{\partial P_0 \over \partial r} -{1 \over \Sigma_0}
{\partial \delta P \over \partial r} +2\Omega \delta u_\phi
-{\partial \delta \Phi \over \partial r}
+a_r~,
\nonumber \\
{d(r \delta u_\phi)\over dt}&=&-{d(r^2\Omega) \over dt}=
-\delta u_r{\partial (r^2\Omega)\over \partial r}~.
\end{eqnarray} 
We assume $P \propto \Sigma^\Gamma$ with $\Gamma$
the adiabatic index. 
    Thus
$\delta P = c_g^2 \delta \Sigma$ 
with $c_g =(\Gamma P/\Sigma)^{1/2}$ the sound speed
in the disk assumed independent of $r$.
Also, we assume  short radial wavelengths
with $|\partial \delta P/\partial r| \gg |\delta P|/r$ and
verify this later.   The continuity equation gives
$\partial (\delta \Sigma)/\partial t 
= -\Sigma_0(\partial \delta u_r/\partial r)$.
The equation for the gravitational potential for $z=0$
gives $|k_r|\delta \Phi = -2\pi G \delta \Sigma$,
where we have taken the perturbations to
be proportional to $\exp(ik_r r)$.

    We now multiply the equation 
for $\delta u_r$ by $(r/\Omega)$ and apply $d/dt$
to it.   Omitting the terms nonlinear in $\delta u_r$,
this gives
\begin{eqnarray}
{\partial^2 \delta u_r \over \partial t^2}&=
&-\left(-c_g^2{\partial^2 \over \partial r^2} + 2\pi i s G \Sigma_0 T
{\partial \over \partial r} +\Omega_r^2\right) \delta u_r 
\nonumber \\
&&+{\partial a_r \over \partial t}
+{\Omega \delta u_r \over r}{\partial\over \partial r} 
\left({r a_r\over \Omega}\right)~.
\end{eqnarray}
Here, $\Omega_r =[r^{-3} d(r^4 \Omega^2)/dr]^{1/2}$
is the radial epicyclic frequency, $s={\rm sign}(k_r)$, and
$T\leq 1$ is a  ``reduction factor'' which is unity
for  $|k_r z_0|<1$ and decreases as $|k_r z_0|$ increases
from unity where $z_0$ is the half-thickness of the
disk (see Shu 1984).

    With $a_r=0$, $\Omega_r=$const, and $\delta u_r
\propto \exp(i k_r r-i\omega t)$ in
equation (15), we recover the dispersion relation for
axisymmetric perturbations of the disk, 
$\omega^2=\Omega_r^2 +(k_r c_g)^2
-2\pi G\Sigma_0 T |k_r|$ (Toomre 1964).

    With $a_r$  given by equation (10) and included
in equation (15), there are {\it forced oscillations}
of the disk due to
the $\partial a_r/\partial t$ term
at radii where $\Omega_r \approx 2\Omega_{bhz}$ (for
the $g_1$ term in equation 11), 
$\Omega_r \approx 4\Omega_{bhz}$ (for the
$g_2$ term), etc.   The radii where these resonances
occur are shown in Figure 5.

     There can be a
 {\it parametric instability} of the disk due to the
$(\Omega/r)\delta u_r\partial (r a_r/\Omega)/\partial r$ term at radii
where $\Omega_r  \approx \Omega_{bhz}$ (for the $g_1$ term),
$\Omega_r \approx 2\Omega_{bhz}$ (for the $g_2$ term), etc.   
  We first discuss the case of forced oscillations.

\begin{figure}
\centering
\includegraphics[scale=0.45]{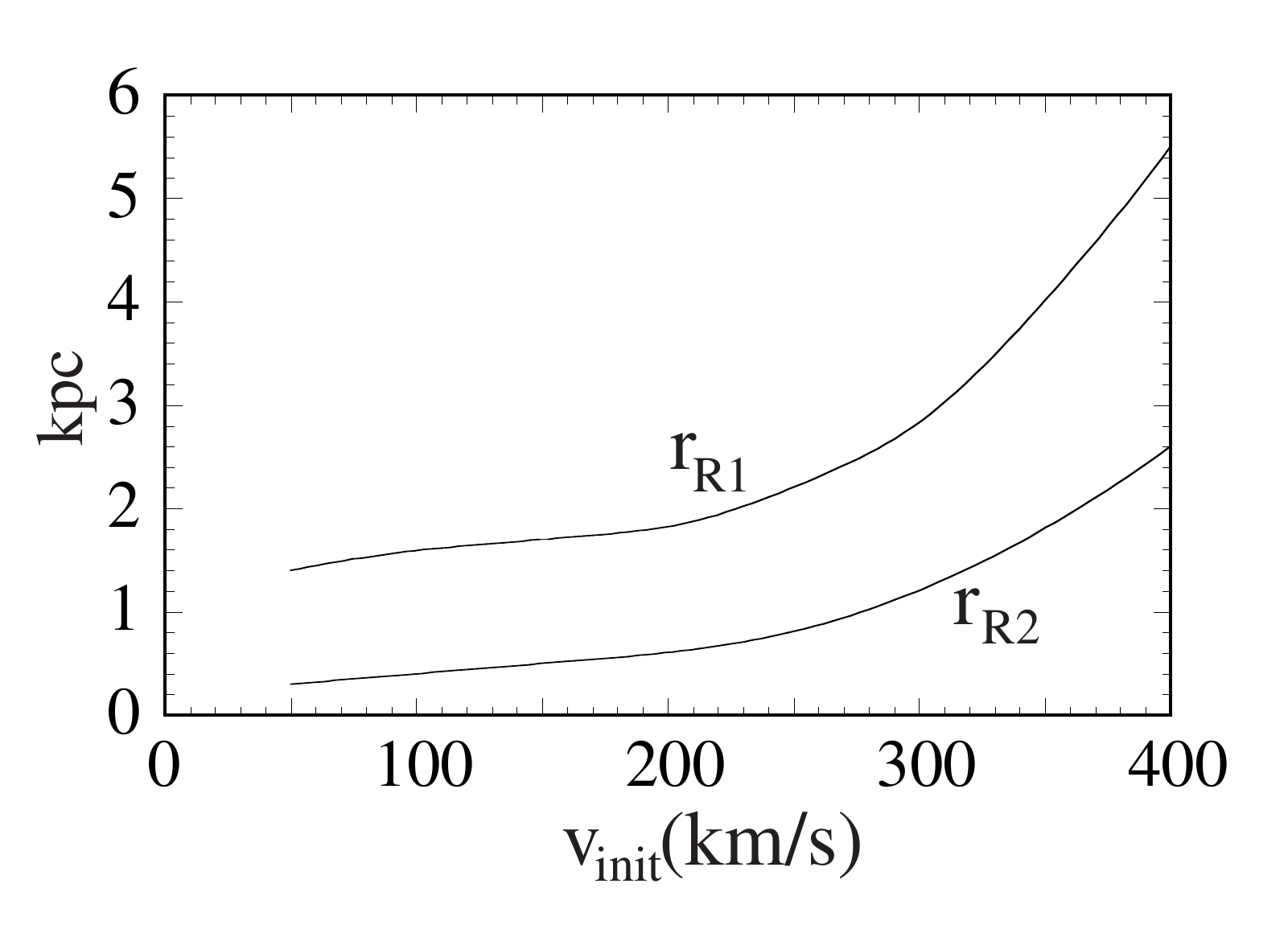}
\caption{Resonant radii where 
$\Omega_r(r_{R1})=2\Omega_{bhz}$ and
$\Omega_r(r_{R2})=4\Omega_{bhz}$ as
a function of the initial BH velocity for $M_{bh}=10^8 M_\odot$.}
\end{figure}

  \subsection{Forced Oscillations of the Disk}

    In the case of forced oscillations  in
equation (15)  we have
$\partial a_r/\partial t=2n\Omega_{bh}^2 r g_n\Omega_{bhz}
\sin(2n\Omega_{bhz}t)$ for $n=1,2,..$ in equation (11).
    At the resonant radius $r_R$ in the disk,
$\Omega_r(r_R)= 2n\Omega_{bhz}$.   
Note that $\Omega_r $ varies slowly across the
disk so that we can write 
$\Omega_r^2=\Omega_{rR}^2[1-2|\eta|(r-r_R)/r_R]$,
where $\Omega_{rR}=\Omega_r(r_R)$ and
$\eta \equiv (r/\Omega_r)(d\Omega_r/dr)$ which
is found to negative for the considered conditions.
We let $\delta u_r (r,t)=\delta u_r(r) \sin(2n\Omega_{bhz}t)$
so that equation (15) becomes
\begin{equation}
\left( {d^2\over dx^2}-ia{d\over dx} +b x\right)
{\delta u_r\over c_g}= {\cal K}(x)~,
\end{equation}
where 
$$
x\equiv {r-r_R \over c_g/\Omega_0}~,\quad
a\equiv {2\pi G\Sigma_0 T\over c_g\Omega_0}~,\quad
b\equiv 2\left({\Omega_{rR}\over \Omega_0}\right)^2|\eta|
\left({c_g\over u_0}\right),
$$
$$
\Omega_0\equiv \Omega(r_R)~,\quad u_0 \equiv r_R \Omega_0~,\quad
{\cal K}= -g_n\left({\Omega_{bh}^2 \Omega_{rR} \over \Omega_0^3}\right)
\left({u_0\over c_g}\right).
$$ 
    Note that for the considered 
thin disks $c_g/u_0 \ll 1$ so that
the $x-$coordinate is an expanded version of $r-r_R$.
    For the assumed conditions we find
$T=1$.
 We can  write $a=2(\Omega_{rR}/\Omega_0)(T/Q)$,
 where $Q=\Omega_{rR} c_g/(\pi G \Sigma_0)$ is Toomre's (1964)
 factor with $Q>1$ disks being stable to axisymmetric
 perturbations.  
 
     We can simplify equation (16) by letting $\delta u_r/c_s=
U(x)F(x)$ and choosing $F=\exp(ia x/2)$.  Then,
\begin{equation}
{d^2 U\over dx^2}+\left({a^2 \over 4}+b x\right)U =\exp(-iax/2){\cal K}(x) ~.
\end{equation}
A WKBJ solution of the homogenous part of the equation
with $U \propto \exp[i\int^x  q(x^\prime)dx^\prime]$ gives
$q=(a^2/4 + b x)^{1/2}$.    Thus there is wave-like
propagation for $x>-a^2/4b$ which begins inside the 
resonant radius where $x=0$.    The region $x<-a^2/4b$
is forbidden and we assume $U(x\rightarrow -\infty)
\rightarrow 0$.
 
 The exact inhomogeneous solution to equation (16) is
 \begin{eqnarray}
{ \delta u_r\over c_g}
 ={\pi e^{iax/2}\over \beta}\big[
 {\rm Ai}( -\alpha-\beta x)\int_{x_\ell}^x dy 
 {\rm Bi}(-\alpha-\beta y){\cal K}(x)e^{-iay/2}
\nonumber \\
-{\rm Bi}(-\alpha-\beta x)\int_{x_\ell}^x dy
{\rm Ai}( -\alpha-\beta y){\cal K}(x)e^{-iay/2}
\big]~,\quad\quad\quad
\end{eqnarray}
where $ \alpha \equiv {a^2 /( 4 b^{2/3})}$, $\beta \equiv b^{1/3}$,
and ${\rm Ai}$ \& ${\rm Bi}$ are the usual Airy functions.
   The value of $x_\ell$ is chosen to be well inside the
forbidden region;  that is, $x_\ell \ll -a^2/4b$.   

\begin{figure}
\centering
\includegraphics[scale=0.45]{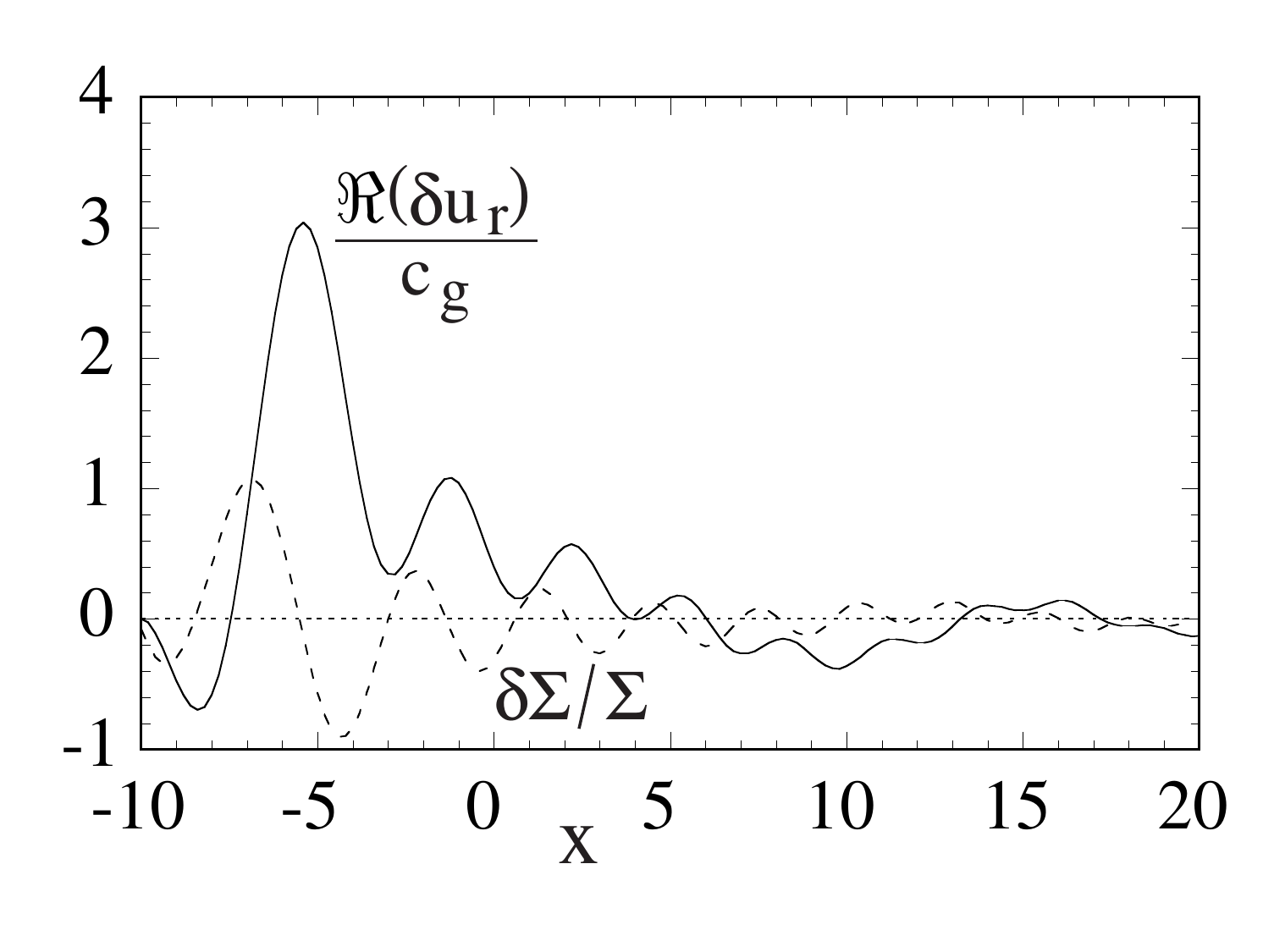}
\caption{Illustrative solution of equation (16) for
$\Re( \delta u_r)/c_g$ (solid curve) and the
corresponding
$\delta \Sigma/\Sigma_0$ (dashed curve)   
for conditions described in the text.  
  For this case $x=20(r-r_R)/r_R$.
Note that
$\delta u_r(x,t)=\Re[\delta u_r(x)]\sin(\Omega_{rR}t)$
and $\delta \Sigma(x,t)=\delta \Sigma(x)\cos(\Omega_{rR}t)$
with $\delta \Sigma(x)/\Sigma_0 =(\Omega_0/\Omega_{rR})
d[ \Re(\delta u_r(x)/c_g)]/dx$.
The location of the
peak of $\delta u_r$ at $x\approx -5.4$ coincides 
approximately with the location of the maximum of
the Airy function ${\rm Ai}(-\alpha-\beta x)$.
}
\end{figure}

Figure 6 shows an illustrative solution of equation (16)
for $\delta u_r/c_g$ and the corresponding 
fractional surface density variations $\delta \Sigma/\Sigma$.
  For this figure we consider
a BH with $M_{bh}=10^8M_\odot$, a resonant radius
$r_R=0.5$ kpc where $\Omega_r(r_R) = 4 \Omega_{bhz}$
(that is, $n=2$).  
  This corresponds approximately to the
results of KL08 with $v_{\rm init}\approx 150$ km/s and
$z_m \approx 1.3$ kpc.  
From Figure 4 we find $g_2 \approx 0.16$.
   For the galaxy model of \S 2.1 we find $f_{bhz}=\Omega_{bhz}/2\pi
\approx 16.9$ Gyr$^{-1}$ (or a period of $59.2$ Myr),
$2\pi/\Omega_{rR} \approx 14.8$ Myr,
$\Omega_{rR}/\Omega_0 \approx 1.89$, and
$\eta \approx -0.311$.   
   We assume the galaxy disk of \S 2.1 has a gas mass-fraction
of $0.22$ and  that $c_g/u_0 =0.05$ so that
the  Toomre factor of the gas disk, $Q=2.1$ which
corresponds to the disk being stable to axisymmetric perturbations.
    For the \S 2.1 model, $u_0=r_R \Omega(r_R) = 110$ km s$^{-1}$
so that $c_g=5.5$ km s$^{-1}$.
We assume $x_\ell =-10.$
    For these values we find $a=1.8$, $b=0.111$, $\alpha=3.52$, 
$\beta =0.480$, and ${\cal K}\approx 
-0.438[1+x(c_g/r_R\Omega_0)]^{-3}$.

    The full-width of the  surface-density peak in
Figure 6    at half-maximum is $\Delta x\approx 2$.
In terms of the actual radius this translates to
$\Delta r \approx r_R(c_g/u_0)^{2/3} K$, where
$K\approx (2\Omega_0/\Omega_{rR})^{2/3}  |\eta|^{-1/3} \approx 1.53$.
We find $\Delta r \approx 0.1$ kpc.
   This length is larger than the ``Toomre-length''  $k_T^{-1} =
r_R(c_g/u_0)(\Omega_0/\Omega_{rR})Q$,  where $k_T$
is the ``least stable wavenumber'' (where $\omega(k_r)^2$ is
a minimum)  for axisymmetric perturbations.   For the
conditions of Figure 6, $k_T^{-1} \approx 0.028$ kpc.

      The magnitude of the disk response, $\delta u_r/c_g$
and $\delta \Sigma/\Sigma_0$,      
to the oscillating BH is directly proportional to the 
BH mass $M_{bh}$ and
inversely proportional to the $Q$-factor of the gas disk. 
   The response decreases strongly as the resonant radius $r_R$ 
increases, roughly as $r_R^{-3}$.   

    A sufficiently strong disk response can lead to shock
formation.   The time-scale for a shock to form is of the
order of $t_{\rm sh} =|\partial \delta u_r/\partial r|^{-1}$.
If $t_{\rm sh}$ is less than half the period of oscillation of
$\delta u_r$ , $\pi/\Omega_r$, then a shock has time to
form.   Using the continuity equation, the condition for
a shock to form is $|\delta\Sigma|/\Sigma_0 > \pi^{-1}$.    
This condition is satisfied for a range of $r$ for the
conditions of Figure 6.   The compression of gas in
the shock wave may in turn lead the formation of
a ring of stars.

     The amplitude of oscillation of the BH, $z_m$, decays
on a time-scale longer than its oscillation period (KL08) due
to dynamical friction of the BH
with the different components of
the galaxy.  Part of the decrease in the BH
kinetic energy goes into the energy of the disk  perturbation.
The decrease of $z_m$ leads to an increase in the BH
oscillation frequency $\Omega_{bhz}$ (KL08).  
Thus the radius of a given resonance $r_R$ decreases
with time.  
    For the case of Figure 6, we estimate that shock formation 
occurs only for $r_R \lesssim 1$ kpc, but continues down
to $r_R \sim 0.2$ kpc.

\subsection{Parametric Instability of the Disk}

      Here we discuss briefly the influence of
 the term $(\Omega/r)\delta u_r
\partial (r a_r/\Omega)/\partial r$ in equation (15) which
   can give rise to a {\it parametric
instability} in the disk.    
   This term can be written as 
   $G_n\Omega_r^2 \delta u_r \cos(2n\Omega_{bhz} t)$,
   for $n=1,~2,..$,
where $G_n = - (\Omega/r)\Omega_r^{-2}
\partial(r^2\Omega_{bh}^2 g_n/\Omega)
/\partial r$ is a dimensionless factor.    
The dominant parametric instability occurs
when $n\Omega_{bhz} \approx \Omega_r$.    
Omitting for simplicity the $r-$derivative terms and
the $\partial a_r/\partial t$  term
in equation (15) gives
\begin{equation}
{\partial^2 \delta u_r \over \partial t^2}+
\Omega_{rR}^2\big\{1-G_n \cos[2(\Omega_{rR}+\delta \Omega_r)t]\big\}
\delta u_r =0~,
\end{equation}
where in this case the ``resonant radius'' $r_{R}$ is  
such that $n\Omega_{bhz} =\Omega_r(r_R)$, and
$\delta \Omega_r =\eta \Omega_r \delta r/r_R$  is a measure of the radial 
distance from $r_R$.
    The solution of this equation is oscillatory with
angular frequency $\Omega_{rR}+\delta \Omega_r$ with
an amplitude exponentially     growing with
growth rate
\begin{equation}
\omega_i:={1\over 4} \Omega_{rR}
\left[G_n^2 -\left({4\delta\Omega_r\over\Omega_{rR}}\right)^2\right]^{1/2}~,
\end{equation}
for $\delta \Omega_r \leq \Omega_{rR}|G_n|/4$ (Landau
\& Lifshitz 1960).  
      The maximum growth rate
is max($\omega_i$)=$\Omega_{rR}|G_n|/4$.
  For the galaxy model of \S 2.1 and a BH mass
$M_{bh}=10^8M_\odot$, we find that $|G_n|$ decreases
from roughly $0.15$ at $r_{rR}=0.2$ kpc to $10^{-3}$ at  $r_{rR}=
1.5$ kpc assuming $g_n=0,16$.  
    We conclude that the parametric instability is 
unimportant compared with the forced motion (\S 2.3) for
$r_{rR} > 0.2$ kpc.

\section{BH Ejection Parallel to the Plane of the Galaxy}

   For the case where the BH is eject parallel to the 
plane of the galaxy we can assume that
the motion is in the $x-$direction,
\begin{equation}
x_{bh}=x_m \sin(\Omega_{bhr} t)~,
\end{equation}   
with $x_m$ the amplitude of the motion
and $\Omega_{bhr}$ the angular frequency
as determined by KL08.   
     The motion of the BH through the gas disk
 of the galaxy is highly supersonic.   
    Thus the BH will leave behind it a narrow
conical shock wave or wake.  
   This shock wave
may trigger star formation particularly in 
regions of high density  such as dense molecular
clouds.
    When the BH is at say $x_{bh}(t^\prime)$ it
leaves a ``shock-section'' at the radius $|x_{bh}(t^\prime)|$ .
At a later time $t\geq t^\prime$ this shock-section   rotates
about the galaxy's centre by an amount
$\Delta \phi = \Omega(|x_{bh}|)(t-t^\prime) \geq 0$.
following the rotation curve of the galaxy.
   Figure 7 shows the geometry of the wake for the 
case where the BH has gone through one
period of radial oscillation, $2\pi/\Omega_{bhr}$.  
   For the case shown this period is $59$ Myr.
   The differential rotation of the disk stretches the wake into
ring-like segments of radii approximately equal to the maximum 
excursion of the BH, $x_m$, where the BH most slowly.

     The relative velocity of the BH and galactic gas
disk is $\Delta {\bf v} = \hat{\rvecphi~} u_\phi(r) -\hat{\bf x}
d x_{bh}/dt$ , where $u_\phi$ is the rotation velocity of
the disk.   As mentioned the flow is highly supersonic
with $|\Delta {\bf v}| \gg c_g$, where $c_g$ is the
sound speed in the gas.       
    In this limit
there is Bondi, Hoyle, \& Lyttleton accretion to the
moving BH (Hoyle \& Lyttleton 1939; Bondi \& Hoyle 1944)
where gas approaching the BH with an impact parameter less than
\begin{equation}
b_{a}={2GM_{bh}\over |\Delta {\bf v}|^2}~,
\end{equation}
accretes to the BH if $b_{a} \leq h$ with $h$ the half-thickness
of the disk.    In this case the cross-section for accretion is
$\pi b_{a}^2$.    If $b_{a} > h$ the cross-section is
approximately $4 b_{a} h$.   We assume $b_a < |x_{bh}|$
during most of the BH's orbit.
    The relevant case is $b_{a}>h$ and this gives the an accretion 
rate    $\dot{M}_{bh} = 4 b_{a}h \rho_g|\Delta {\bf v}|$ where
$\rho_g$ is the gas density. 
   The accretion time-scale is then
\begin{eqnarray}
T_{a}\!\!\!\!\!&=&\!\!\!\!\!{M_{bh} \over \dot{M}_{bh}}=
{|\Delta {\bf v}| \over 8 G h \rho_g}~,
\nonumber \\
&\approx&\!\!\! \!\!230{\rm Myr}
\left({50{\rm pc} \over h}\right)
\left({10{\rm H cm}^{-3}\over \rho_g}\right)
\left({|\Delta {\bf v}|\over 100~\!{\rm km/s}}\right).
\end{eqnarray}
For these reference values $b_{a}\approx 87$ pc and
$\dot{M}_{bh}\approx 0.43 M_\odot$ yr$^{-1}$.
   The drag force on the BH is simply
${\bf F}_{bh} = 4 \rho_g b_{a}h|\Delta{\bf v}|
\Delta {\bf v}$.  This acts to both slow down the BH
radial motion and impart to it azimuthal motion in
the rotation direction of the galaxy.  
    The time-scales for the changes in the motion are
all of order $T_{a}$.

\begin{figure}
\centering
\includegraphics[scale=0.45]{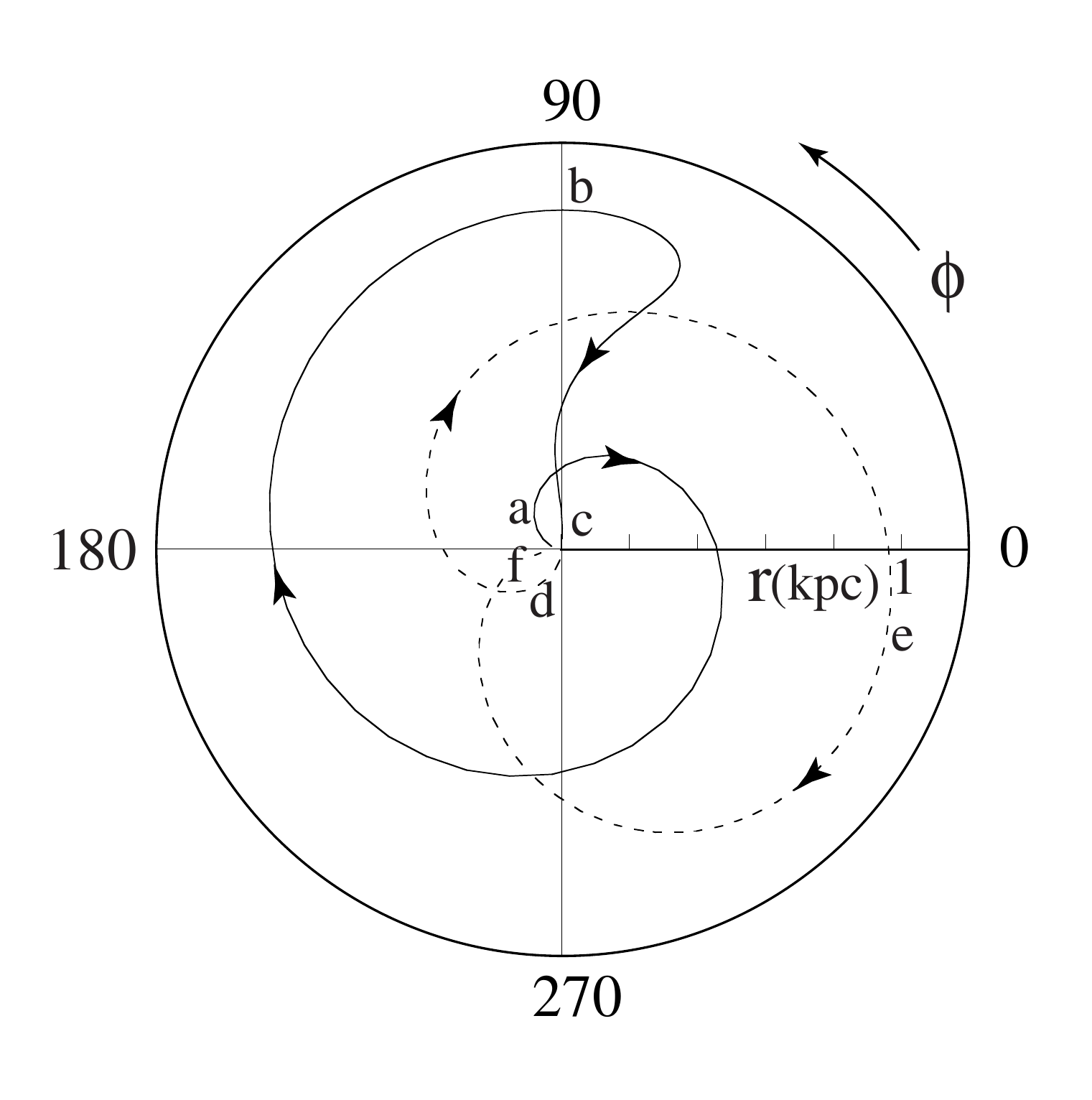}
\caption{Wake of the BH ejected in the  plane of the
galaxy for the case where the BH has gone through
one period of radial oscillation, $2\pi/\Omega_{bhr}$.   
  For this plot $x_m=1$ kpc and $f_{bhr}=\Omega_{bhr}/2\pi
=17$ Gyr$^{-1}$ corresponding to a period of $59$ Myr.  
The points marked by $a,b, ..$ correspond to the
$t^\prime =0$, $t^\prime =\pi/(2\Omega_{bhr})$,
etc.
}
\end{figure}

\section{Conclusions}
    
    Here, we first analyzed the linear axisymmetric
perturbations of a gas disk driven by a black hole oscillating vertically
along the axis of symmetry. 
       We find 
that there is a strong resonant  forcing
of the disk radial motion near ``resonant radii'' $r_R$ where the BH oscillation frequency is equal one-half, one-fourth, ($1/6$, etc.) of the radial
epicyclic frequency in the plane of the disk.
    Near the resonant radii
there can be a strong enhancement of the radial flow velocity and
disk density  which can lead to shock formation. 
   The shock formation occurs during one period of oscillation
of the BH which is assumed longer than the BH damping
time due to dynamical friction.
      This shock may trigger the
formation of a ring of stars near $r_R$.
    As an example, for a BH mass of $10^8~M_\odot$ and a kick
velocity of $150$ km s$^{-1}$,    
we find that the resonant radii lie between $0.2$ and $1$ kpc.
     The magnitude of the disk response is  proportional
to the BH mass, inversely proportional to the Toomre~$Q$ of the gas disk,
and decreases rapidly as $r_R$ increases.  
   The resonant radii increase as the initial BH kick
velocity increases (KL08).

      For BH motion parallel to the plane of the galaxy
we find that the BH leaves behind it a supersonic wake which
over time gets contorted into a complicated shape by the galaxy's
differential rotation.   The shape  of the
wake is calculated for an illustrative case  
as well as the slow-down time of the BH.
     The differential rotation of the disk stretches the wake into
ring-like segments of radii approximately equal to the maximum 
excursion of the BH, $x_m$, where the BH most slowly.  

   Many other processes may be involved in the
formation and evolution of nuclear rings in
galaxies (e.g.,  van de Ven \& Chang 2009).

\section*{Acknowledgements}

We thank Drs. L.E. Kidder, M.S. Tiscareno, M.M. Hedman,
and Profs. D. Lai and R. Giovanelli for helpful discussions.
This work has made use of the computational facilities of the National
Astronomy and Ionosphere Center, which is operated by
Cornell University under a cooperative agreement with the National
Science Foundation.
RVEL  was supported in 
part by NASA grant NNX08AH25G and by
NSF grants AST-0607135 and AST-0807129.

\end{document}